\documentclass[final,3p,twocolumn]{elsarticle}
\usepackage{amsmath}
\usepackage{graphicx}
\usepackage{amssymb}
\usepackage{times}

\begin{document}

\title{Fast Afterglows of Fast Radio Bursts}

\author{Andrei Gruzinov}

\begin{abstract}
The main FRB event may leave behind a clump of relativistic plasma with high ``free energy'' density. As the plasma undergoes collisionless relaxation, it emits coherent electromagnetic waves. These electromagnetic waves may be observable as a fast radio afterglow, with decreasing frequency and intensity. We demonstrate the fast coherent afterglow in a numerical experiment. We tentatively predict the peak afterglow frequency decreasing with time as $\nu \propto t^{-3/2}$. 

\end{abstract}

\begin{keyword}
Radio transient sources
\end{keyword}

\maketitle

\section{FRB origin and mechanism}
\label{sec:intro}

The {\it place of origin} of FRBs can perhaps be established by observations in the near future. E.g., \cite{Kirs} ``...conclusively prove that FRB 20200120E is associated with a globular cluster...'' and claim that ``...this association challenges FRB models that invoke magnetars ...FRB 20200120E is a young, highly magnetised neutron star..''

Unlike the {\it place of origin}, the {\it physical mechanism(s)} of FRBs will probably remain uncertain for longer. This is because FRBs are likely produced by relativistic plasma, and any relativistic plasma with high ``free energy'' density naturally emits or, more precisely, contains electromagnetic waves. The electromagnetic waves are excited together with various other plasma waves; the waves are just available degrees of freedom, nothing remarkable. If so, many different plasmas can make an FRB. How can we find out which plasmas do actually make FRBs?

Very short FRBs are the best hope for understanding the {\it physical mechanism(s)}. Suppose one detects an FRB with individual pulses even shorter than the 100 ns ones of \cite{Maji}. Say, just a few tens of wavelengths. For such ultra-short FRBs, a faithful numerical simulation must be doable.  Then we can: (i) make a list of physical setups which theoretically should give an FRB, in order of increasing sophistication; (ii) numerically simulate all setups at the top of the list; (iii) see if some of the numerical results agree with observations. This approach will work iff actual FRBs are simple and our list of simple physical setups is full enough. 

The absolute champion of simplicity among physical setups leading to an FRB is the Simple Shmaser (SHock MASER, \cite{Gruz}). We briefly describe the Simple Shmaser in \S\ref{sec:shma}. While running Simple Shmaser experiments (numerical, using simplified, non-faithful numerics, as described in \S\ref{sec:PIC}), we observed the fast afterglow phenomenon. The afterglow must be a generic feature of many other FRB setups, and one can even make a (very simple) prediction regarding the time dependence of the fast afterglow peak frequency, \S\ref{sec:glow}. 

Fast FRB afterglows, if observed, will help constrain theoretical models. Even more interesting would be a non-detection of fast FRB afterglows, with strong upper bounds -- many theoretically possible physical setups would become observationally excluded. 

~

~

\section{Simple Shmaser}\label{sec:shma}

Masing (coherent emission of electromagnetic waves) in a magnetized relativistic plasma is simple in principle, but very complicated in full detail \citep{Gruw}. No theory of masers should be trusted without confirmation by direct numerical experiments.

The simplest physical setup leading to a maser action is the Simple Shmaser \citep{Gruz}. Collide two cold (and therefore unmagnetized) electron-positron plasma clouds. Assume that in the CM frame the clouds have similar sizes, $R$, similar densities $n$, and that the clouds collide at a mildly relativistic relative velocity. Further assume that the CM frame is boosted by a high Lorentz factor, $\Gamma$. \cite{Gruz} claims (and confirms for very short FRBs) that the resulting FRB will have the duration, $\tau$, the peak frequency, $\nu$, and the equivalent isotropic energy, $E$, given by
\begin{equation}\label{eq:t1}
\tau \sim \Gamma ^{-1}\frac{R}{c},
\end{equation}
\begin{equation}\label{eq:t2}
\nu \sim \Gamma \left(\frac{ne^2}{\pi m}\right) ^{1/2},
\end{equation}
\begin{equation}\label{eq:t3}
E\sim 0.01\Gamma ^3mc^2R^3n,
\end{equation}
where $m$ is the electron mass. In words, the radio waves are emitted as long as the clouds keep colliding (giving $\tau$), at about the plasma frequency (giving $\nu$), with $\sim 0.01$ radiative efficiency (giving $E$). The powers of $\Gamma$ are from relativistic kinematics. The $\sim 0.01$ efficiency was taken from (non-faithful, because two-dimensional) numerical simulations. For a one-dimensional Simple Shmaser, the equations of motion are written in the Appendix, where the shock-mediating instabilities are calculated.

For any set of FRB parameters $(\tau,\nu,E)$, one can find a set of the Simple Shmaser parameters $(R,n,\Gamma)$ which gives rise to the desired FRB (we must repeat: we have only two-dimensional and one-dimensional simulations, and only for very short FRBs, lasting for up to a few tens of wavelengths). 

That three observed parameters can be replaced by three theoretical parameters, obviously, proves nothing. But one can start experimenting with the Simple Shmaser, and see what comes out of it. We saw an afterglow, as described, and then generalized, below. 

~

\section{Simple Shmaser in one dimension: Numerical Experiment }\label{sec:PIC} 

One wants to look at many FRB experiments, at high resolution, hoping to see some regularities in the mostly chaotic radio emission. For this author, high resolution plus many numerical simulations equals one dimension. The one-dimensional Simple Shmaser is:
\begin{itemize}
\item collisionless electron-positron plasma
\item the distribution functions of electrons and positrons depend on time, $t$, one spatial dimension, $x$, and two 4-velocity components, $u_x$, $u_y$. The particles are not moving along $z$, $u_z=0$.
\item the electromagnetic field has the following components: $E_x(t,x)$, $E_y(t,x)$, $B_z(t,x)$.
\end{itemize}
The equations of motion are given in the Appendix, where we also describe relevant linear instabilities.

\begin{figure}
\includegraphics[width=0.5\textwidth]{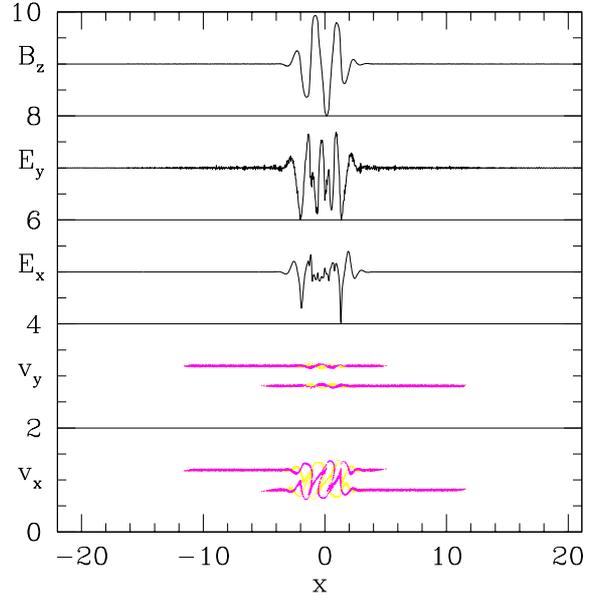}
\caption{In natural units, at $t=23.7$, vs $x$, shown are $1+v_x$, $3+v_y$ (for 20000 randomly selected particles, electrons and positrons are shown by different colors), $5+E_x/0.189$, $7+E_y/0.00484$, $9+B_z/0.0780$}
\label{f1}
\end{figure}

We solve Maxwell-Lorentz equations of motion for 1000000 particles, using 20000 grid points for the fields. Initially, we have two constant density, $n$, clouds moving towards each other, with 4-velocities $u_x=\pm 0.2$, $u_y=\pm 0.2$, see Fig.(\ref{f1}). Since collisionless plasma ($\equiv$Vlasov$\equiv$collisionless Boltzmann+Maxwell-Lorentz) only knows about the charge and mass densities, but not about the charge and mass of any individual particle, the (rationalized) ``natural units'' of the problem are 
\begin{equation}
ne=nm=c=1.
\end{equation}
We use these units in what follows: the time unit is then the inverse plasma frequency (calculated from the CM density of the clouds before the collision), the length unit is the skin depth, the electromagnetic field unit is $(nmc^2)^{1/2}$.

As shown in Fig.(\ref{f1}), with initial cloud sizes equal to 15.8, by $t=23.7$, the two-stream instability fully develops in the region of clouds interpenetration. At this time, the field is mostly electrostatic -- $E_x$ dominates. Next in order is the magnetostatic field $B_z$ generated by the Weibel instability. What little electromagnetic waves ($E_y$-$B_z$ mixture) we have, are fully confined within the plasma; the ultimate fate of the electromagnetic waves is uncertain at this time.

\begin{figure}
\includegraphics[width=0.5\textwidth]{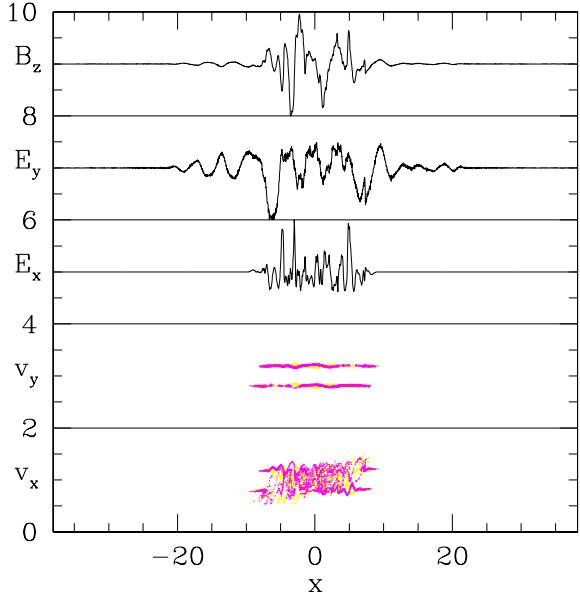}
\caption{$t=43.3$, $5+E_x/0.122$, $7+E_y/0.00801$, $9+B_z/0.0426$}
\label{f2}
\end{figure}

However, as shown in Fig.(\ref{f2}), by $t=43.3$, the electromagnetic waves do manage to escape from the plasma. Now the plasma can never catch up with the outgoing waves, and therefore we have a successful FRB. No surprises here, we have already seen a similar proof of principle -- Simple Shmasers give FRBs -- in a more faithful two-dimensional simulation of \cite{Gruz}. But in one dimension we can keep running the Simple Shmaser experiment for a much longer time...

\section{Fast Afterglows of FRBs}\label{sec:glow}

\begin{figure}
\includegraphics[width=0.5\textwidth]{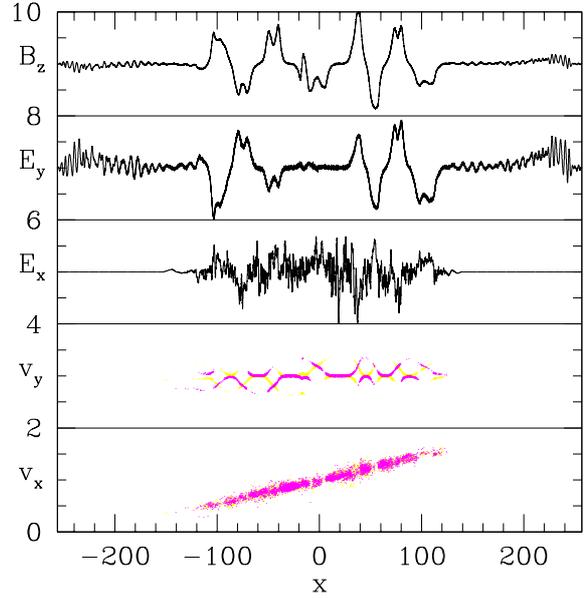}
\caption{$t=278$, $5+E_x/0.0087$, $7+E_y/0.0094$, $9+B_z/0.031$}
\label{f3}
\end{figure}

Fig.(\ref{f3}) shows the particles and fields long after the collision, at $t=278$. As expected from the ``theory'' Eqs.(\ref{eq:t1}-\ref{eq:t3}): 
\begin{itemize}
\item The characteristic emitted wavelength is $\sim 2\pi$, because the plasma frequency is $\omega_p=1$ in natural units.
\item The main FRB event (high-amplitude wavepackets at $|x|\approx 240$), contains a few waves, $=$ the initial cloud size divided by $2\pi$. The number of waves in the main FRB event roughly matches the number of waves seen the $(x,v_y$) projection of the particle phase space. 
\item The measured radiative efficiency, $\equiv$ the fraction of the clouds' kinetic energy converted to outgoing electromagnetic waves, is $\sim 0.001$ at $t=278$, and is not expected to grow substantially. This is a factor of 10 below the $0.01$ efficiency used in Eq.(\ref{eq:t3}) --  still ``as expected from the theory'', as our ``theory'' is but a rough estimate. 
\end{itemize}

As shown in Fig.(\ref{f3}), even at a very late time $t=278$, when the shock waves are long gone, the plasma keeps emitting electromagnetic waves. This is the coherent afterglow stage. As seen at $x\approx150$, the afterglow frequency is half the main FRB frequency, the intensity is about 0.1 of the main FRB intensity, at a time delay $\approx 5$ main FRB durations, for the main FRB that lasted  $\approx 5$ wavelengths.

Now we must try to guess, assuming Simple Shmaser, what will happen in 3 dimensions. We must predict some characteristic of the afterglow. Then, if observers do see it, we will have a (weak, indirect) confirmation of the Simple Shmaser model. We will ``predict'' the time dependence of the peak frequency of the afterglow.  

The only trustworthy way to predict anything about FRBs is numerical simulations in 3 dimensions, as we discuss at length in \cite{Gruz}. This author can't do long enough 3-dimensional Vlasov simulations. But there exists a meaningful lower bound on the frequency; we will assume that the lower bound and the peak frequency have the same time dependence. 

The electromagnetic wave dispersion law, assuming the plasma is non-relativistic is $\omega^2=k^2+\omega_p^2$,~~ $\omega_p^2\equiv \frac{4\pi ne^2}{m}$. The Simple Shmaser plasma is only mildly relativistic, and this dispersion law is roughly valid. The electromagnetic waves propagating in the plasma must have high enough frequency, $\omega > \omega_p$. As the plasma clump expands in three dimensions, the density drops as $n\propto t^{-3}$. Since $\omega_p^2\propto n$, we get the following peak frequency estimate\begin{equation}
\nu \propto t^{-3/2}.
\end{equation}

This prediction is just an illustration, a mere possibility. In different scenarios, electrons in the shocked plasma can be ultra-relativistic, with characteristic Lorentz factor $\gamma \gg 1$. Then $\omega_p^2\approx \frac{4\pi ne^2}{\gamma m}$ (in a sense, see, e.g., \cite{Gruw} Eq.(27) and correct their Eq.(30) to $\epsilon_\perp\omega^2=k^2$). With $\gamma \propto t^{-1}$, we now get
\begin{equation}
\nu \propto t^{-1}.
\end{equation}

~

\section{Conclusion}\label{sec:conc}

We tentatively predict fast coherent afterglows of Fast Radio Bursts. The peak afterglow frequency decreases as  $\nu \propto t^{-3/2}$ or $\nu \propto t^{-1}$, perhaps. If detected, coherent afterglows may help establish the FRB emission mechanism. Not seeing fast FRB afterglows, with good upper bounds, would be even more interesting -- such an observation would rule out many models.

~

I thank Sterl Phinney for a useful discussion and for telling me about 100ns FRBs. 

~

\appendix

\section{Simple Shmaser in one dimension: equations and linear instabilities}

Consider electromagnetic field of the following special geometry
\begin{equation}
{\bf E}=(E_x(t,x),E_y(t,x),0),~~~{\bf B}=(0,0,B_z(t,x)),
\end{equation}
and positron/electron distribution functions of the form
\begin{equation}
f_{p,e}=f_{p,e}(t,x,p_x,p_y).
\end{equation}

In this ``1d2v'' case, Vlasov equations ($\equiv$ Maxwell-Lorentz plus collisionless Boltzmann) take the form
\begin{equation}\label{eq:v1}
\begin{split}
\partial_tf_{p,e}+v_x\partial_xf_{p,e}\pm e(E_x+B_zv_y)\partial_{p_x}f_{p,e} \\
\pm e(E_y-B_zv_x)\partial_{p_y}f_{p,e}=0,
\end{split}
\end{equation}
\begin{equation}\label{eq:v2}
\partial_xE_x=4\pi e \int d^2p(f_p-f_e),
\end{equation}
\begin{equation}\label{eq:v3}
\partial_tE_y=-\partial_xB_z-4\pi e\int d^2p(f_p-f_e)v_y,
\end{equation}
\begin{equation}\label{eq:v4}
\partial_tB_z=-\partial_xE_y,
\end{equation}
where
\begin{equation}
{\bf p}=m{\bf u},~~~{\bf u}=\gamma{\bf v},~~~\gamma^2=1+u^2.
\end{equation}

All physical states with zero electromagnetic field, $E_x=E_y=B_z=0$, and with equal homogeneous distributions, $f_p=f_e=F({\bf p})$ are equilibria. For some $F({\bf p})$, the equilibria are unstable. Before the clouds interpenetrate, we have $F({\bf p})=n\delta({\bf p}\pm{\bf p}_0)$ in the right/left cloud, $n$ is the initial density.  These states are, of course, stable -- all particles are at rest in the appropriate frame. 

After the clouds interpenetrate, we initially get the distribution $F({\bf p})=n(\delta({\bf p}-{\bf p}_0)+\delta({\bf p}+{\bf p}_0))$. This equilibrium distribution is linearly unstable. We will derive the linear instability growth rates for two limiting cases. The limiting cases clarify the operation of the Simple Shmaser and explain the origin of the theoretical estimates Eqs.(\ref{eq:t1}-\ref{eq:t3}).

\subsection{Two-Stream} 

The two-stream instability occurs because one-dimensional positive (negative) charges traveling through an electrostatic potential well speed up (slow down) and thereby make the well deeper.

Take ${\bf p}_0$ along $x$. Assume $E_y=B_z=0$ and $f_{p,e}\propto \delta(p_y)$ -- the particles keep moving only along $x$. This is a self-consistent assumption -- without either $E_y$ or $B_z$ the particles will keep moving along $x$, there will be no current along $y$, and $E_y$, $B_z$ will remain zero. Then, with the ansatz $\propto e^{-i\omega t+ikx}$, the basic equations (\ref{eq:v3},\ref{eq:v4}) are satisfied trivially, while linearized Eqs.(\ref{eq:v1},\ref{eq:v2}) read
\begin{equation}\label{eq:v11}
-i\omega \delta f_{p,e}+v_xik\delta f_{p,e}\pm eE_x\partial_{p_x}F=0,
\end{equation}
\begin{equation}\label{eq:v21}
ikE_x=4\pi e \int d^2p(\delta f_p-\delta f_e).
\end{equation}

Calculate $\delta f_{p,e}$ from Eq.(\ref{eq:v11}), then Eq.(\ref{eq:v21}) reads
\begin{equation}
k=-8\pi ne^2\int dp_x\frac{1}{\omega-kv_x}\frac{d}{dp_x}(\delta(p_x-p_0)+\delta(p_x+p_0)).
\end{equation}
Integrate by parts:
\begin{equation}
1=\frac{\omega_p^2}{(\omega-kv_0)^2}+\frac{\omega_p^2}{(\omega+kv_0)^2}, ~~~~~~\omega_p^2\equiv \frac{8\pi ne^2}{\gamma_0^3m}.
\end{equation}
The unstable branch (two-stream instability) is
\begin{equation}
\omega^2=\omega_p^2+v_0^2k^2-\omega_p\sqrt{\omega_p^2+4v_0^2k^2}.
\end{equation}
The fastest growing mode is
\begin{equation}
-i\omega=\frac{1}{2}\omega_p,~~~~~~k=\frac{\sqrt{3}}{2}\frac{\omega_p}{v_0}.
\end{equation}
For mildly relativistic cloud collisions, the characteristic time scale, $\sim \omega_p^{-1}$, agrees with our ``theory'' Eq.(\ref{eq:t2}). Of course, the two-stream instability is electrostatic, but the time scale can still matter. Once we go fully nonlinear 3-dimensional, there will be various mode conversions, modulations, etc. 

\subsection{Weibel} 

Weibel instability occurs because like currents attract. 

Take ${\bf p}_0$ along $y$ (strictly speaking we mean nearly along $y$, for the clouds need to interpenetrate). Assume $E_x=0$ and $f_e(t,x,p_x,p_y)=f_p(t,x,p_x,-p_y)$ -- the electrons and positrons move in unison. This is a self-consistent assumption -- without $E_x$, a pair of opposite charges which starts in unison (same $x,v_x$, equal in magnitude but opposite $v_y$) will stay in unison, creating no charge density (as projected on $x$), and therefore no $E_x$.  Then, with the ansatz $\propto e^{-i\omega t+ikx}$, the basic equation (\ref{eq:v2}) is satisfied trivially, while linearized Eqs.(\ref{eq:v1},\ref{eq:v3},\ref{eq:v4}) read
\begin{equation}\label{eq:v12}
\begin{split}
-i\omega \delta f_{p,e}+v_xik\delta f_{p,e}\pm eB_zv_y\partial_{p_x}F \\
\pm e(E_y-B_zv_x)\partial_{p_y}F=0,
\end{split}
\end{equation}
\begin{equation}\label{eq:v32}
-i\omega E_y=-ikB_z-4\pi e\int d^2p(\delta f_p-\delta f_e)v_y,
\end{equation}
\begin{equation}\label{eq:v42}
-i\omega B_z=-ikE_y,
\end{equation}

Now calculate $B_z$ in terms of $E_y$ from Eq.(\ref{eq:v42}), calculate $\delta f_{p,e}$ in terms of $E_y$ from Eq.(\ref{eq:v12}), and get the dispersion law
\begin{equation}
\omega^2=k^2-8\pi e^2\int d^2pv_y\left( \frac{kv_y}{\omega-kv_x}\frac{\partial}{\partial p_x}+\frac{\partial}{\partial p_y} \right) F.
\end{equation}
Integrate by parts, plug in $F=n\delta(p_x)(\delta(p_y-p_0)+\delta(p_y+p_0))$:
\begin{equation}
\omega^2=k^2+\omega_p^2\left(1+\frac{u_0^2k^2}{\omega^2}\right), ~~~~~~\omega_p^2\equiv \frac{16\pi ne^2}{\gamma_0^3m}.
\end{equation}
The unstable branch is
\begin{equation}
\omega^2=\frac{1}{2}\left( \omega_p^2+k^2-\sqrt{(\omega_p^2+k^2)^2+4\omega_p^2u_0^2k^2}\right).
\end{equation}
For $\omega_pu_0k\ll \omega_p^2+k^2$ (and therefore, approximately, for all $k$ in mildly relativistic shocks), the instability growth rate is
\begin{equation}
-i\omega\approx \frac{\omega_pu_0k}{\sqrt{\omega_p^2+k^2}}.
\end{equation}
Although all small scale perturbations (smaller that the skin depth, $\omega_p^{-1}$) grow at about the same rate, numerical work shows that nonlinear Weibel manifests at about the skin depth. So again, the characteristic time scale of the instability agrees with the ``theory'' Eq.(\ref{eq:t2}). 

We can now explain the Simple Shmaser ``theory'' Eqs.(\ref{eq:t1}-\ref{eq:t3}). We will work in the CM frame, $\Gamma=1$, boosting the results to large $\Gamma$ is straightforward. As plasma clouds interpenetrate, the two-stream/Weibel instabilities mediate two shock waves running into the clouds. The instabilities create nonlinear structures with characteristic frequency given by Eq.(\ref{eq:t2}) and must emit electromagnetic waves with the same characteristic frequency. The emission lasts for as long as the shock waves propagate through the clouds, hence Eq.(\ref{eq:t1}). Numerical work shows that instabilities saturate when quasi-electrostatic and quasi-magnetostatic fields take on about 3-10\% of the plasma kinetic energy density. The resulting electromagnetic wave emission into empty space is even less efficient, giving the 0.01 numerical coefficient in Eq.(\ref{eq:t3}).

The major uncertainty of the Simple Shmaser is the $R$-scaling. Yes, collisionless shocks propagate through plasma for a time $\sim \frac{R}{c}$. Yes, electromagnetic waves of frequency $\sim \omega_p$ must be emitted and are seen to be emitted in numerical experiments. But -- do the electromagnetic waves get out of the plasma? Should we not expect non-linear Landau damping?

We know from numerical simulations that nonlinear absorption does not shut down the Simple Shmaser at least for up to a few tens of emitted wavelengths. Until proven otherwise, Simple Shmaser is a viable model for FRBs consisting of $\sim 100$ ns pulses at $\sim 1$ GHz, similar to \cite{Maji}.

\bibliographystyle{hapj}

\end{document}